\documentclass{emulateapj}
\usepackage{natbib}
\usepackage{graphicx}
\usepackage{epstopdf}
\DeclareMathSymbol{\varOmega}{\mathord}{letters}{"0A}
\DeclareMathSymbol{\varSigma}{\mathord}{letters}{"06}

\DeclareMathSymbol{\varPsi}{\mathord}{letters}{"09}

\newcommand{\Eq}[1]{Equation\,(\ref{#1})}
\newcommand{\Eqs}[2]{Equations (\ref{#1}) and~(\ref{#2})}

\newcommand{\Fig}[1]{Figure~\ref{#1}}
\newcommand{\Figs}[2]{Figures~\ref{#1} and \ref{#2}}

\newcommand{\icarus}{\rm{Icarus}}

\newcommand{\ltsim}{\protect\raisebox{-0.5ex}{$\:\stackrel{\textstyle <}{\sim}\:$}}

\newcommand{\Msun}{M_{\odot}} 
 
\newcommand{\Rsun}{R_{\odot}}

\newcommand{\lsim}{\mathrel{\rlap{\lower4pt\hbox{\hskip1pt$\sim$}}
   \raise1pt\hbox{$<$}}}                
\newcommand{\gsim}{\mathrel{\rlap{\lower4pt\hbox{\hskip1pt$\sim$}}
   \raise1pt\hbox{$>$}}}                

\begin{document}

\title{Star Hoppers: Planet Instability and Capture in Evolving Binary Systems}

\author{Kaitlin M. Kratter \& Hagai B. Perets}

\affil{Institute for Theory and Computation, Harvard-Smithsonian Center for Astrophysics, 60 Garden St.; Cambridge, MA, USA 02138}

\begin{abstract}
Many planets are observed in stellar binary systems, and their frequency
may be comparable to that of planetary systems around single
stars. Binary stellar evolution in such systems influences
the dynamical evolution of the resident planets. Here we study the
evolution of a single planet orbiting one star in an evolving binary system. We find
that stellar evolution can trigger dynamical instabilities that drive planets into chaotic orbits. 
This instability leads to planet-star collisions, exchange
of the planet between the binary stars ({}``star-hoppers"), and ejection of the planet from the system. The means by which planets can be recaptured is similar to the pull-down capture mechanism for irregular  solar system satellites.
Because planets often suffer close encounters with the primary on the asymptotic giant branch, captures during a collision with the stellar envelope are also possible. Such capture could populate the habitable zone around white dwarfs.
\end{abstract}

\section{Introduction}
Most of the ($>2000$) planets and planet candidates discovered to date are hosted by  single stars \citep{Howard:2011}. This bias towards single stars may be primarily due to the avoidance of multiple systems in target selection: the limited number of planet discoveries in binary systems suggests that the planet frequency in multiples may be comparable to that of single stars \citep{Eggenberger:2004,Bonavita:2007,Mugrauer:2009}. Given the frequency of binary systems (\citealt{Raghavan2010}), a large fraction of all exoplanetary systems may reside in binaries. Here we focus on the dynamics of planets in evolving binary systems near the peak of the binary separation distribution: from $\sim 50-100$ AU.

Though various studies have explored the possible formation and dynamical evolution of planets in binary systems \citep{Holman:1999,David:2003,Mudryk:2006,Perets:2010,Moeckel:2012}, the effects of stellar evolution on such systems have not been thoroughly explored.  The influence of stellar evolution on planets in singleton systems has been studied by \cite{Debes:2002,Villaver:2009,Veras:2011}. \cite{Perets:2010} described qualitatively the evolution of planets in evolved binary systems, including both circumstellar and circumbinary planets.  \cite{Veras:2012} have recently conducted a quantitative study of circumbinary planetary dynamics in evolving systems.

In this paper we focus on the effects of binary stellar evolution on planetary systems in ``satellite" orbits, those around one star in the system.  We consider the influence of slow mass loss via stellar winds, and demonstrate that these systems have rich dynamical behavior, including planetary ejections, ``star hopping" (exchanges of planets between the binary stars), capture of planets into stable orbits, collisions with either of the stars, and interactions potentially leading to capture into close  orbits. A similar mechanism that operates in triple stellar systems, dubbed the Triple Evolution Dynamical Instability (TEDI)  has been recently explored by \cite{PK2012}. 

In the following we explore the outcomes of this instability using few-body simulations, and provide a qualitative and semi-quantitative understanding of the properties of such systems. We also provide an estimate for the fraction of systems that are subject to the planetary version of the TEDI, given various assumptions for the population of planetary systems in binaries.

We begin in Section \ref{sec-dynamical} by describing the orbital evolution and stability criterion for planets in evolving binaries.  We next explain the capture mechanism for planets around the secondary star in the context of the circular restricted three body problem (Section \ref{sec-jacobi-cap}). In Section \ref{sec-nbody} we describe a series of n-body experiments that we use to explore the outcomes of the chaotic planetary orbits induced by stellar mass loss. In Section \ref{sec-results}  we present probabilities for collision and capture for planets in our numerical experiments,  and provide a summary of expected outcomes.  In Section \ref{sec-frequency} we explore the frequency of unstable planetary binary systems. Finally,  in Section \ref{sec-discussion} we conclude with several observational predictions for next generation surveys.

\section{Dynamical evolution of circumstellar planets in evolving binary systems}  \label{sec-dynamical}    
In order to remain dynamically stable, a triple system of gravitating objects requires a hierarchical configuration with orbits that are well separated.  For satellite type planetary-binary systems, there is a critical ratio between the semi-major axes of the binary and the planet beyond which the system is unstable. The stability boundary is quite complex in detail, set by the overlap of mean motion resonances \citep{Wisdom:1980,Mudryk:2006}, however it can be well fit as a function of binary mass ratio and eccentricity \citep{Holman:1999}. For circular binary orbits, which we consider here, the radius of the stable region around the primary is:
\begin{equation}
R_p =  \left(0.464 - 0.38\frac{m_2}{m_1+m_2}\right) a_* \label{eq:hw-approx}
 \end{equation}
 where $a_*$ is the stellar semi-major axis, and $m_1$ and $m_2$ are the masses of the primary and secondary, respectively. Note that this is identical to Equation (1) of \cite{Holman:1999}, but without the quadratic dependence on eccentricity. 
These authors find that planets with semi-major axis $a_p > R_p$ are destabilized within $10^4$ binary orbital periods, (see also the results of \citealt{David:2003} and \citealt{Fatuzzo:2006}, who calculate typical instability timescales). The size of the stability region around the secondary, $R_s$, is obtained by the substitution $m_1\leftrightarrow m_2$.  Note that the initial eccentricity of the planet, which we set to zero here, is unimportant for stability considerations \citep{Mudryk:2006}.

We now consider the impact of slow (adiabatic) and isotropic mass loss from the primary star in the  binary, where the mass loss timescale  is long compared to the planetary and stellar orbital periods.  Mass loss is the byproduct of stellar evolution and occurs through stellar winds.  Adiabatic mass loss drives the orbits to larger semi-major axes: 
\begin{equation}
a_{f}=\frac{M_{i}}{M_{f}}a_{i},\label{eq:adiabatic}
\end{equation}
\citep{Hadjidemetriou:1963},
where $M_{f}$ is the final mass of the system after its evolution,
and $M_{i}$ and $a_{i}$ are the initial mass and initial semi-major axis
of the system, respectively.  This relation derives from a perturbation analysis that shows that the the specific angular momentum $GMa(1-e^2)$ (per reduced unit mass) is conserved,
and the assumption that mass loss is slow and isotropic, which guarantees that eccentricity is also conserved.
 These assumptions are only accurate for planets far
from their host star (i.e. beyond the tidal effects of interaction with the expanding envelope of the evolving primary), and where mass transfer is negligible. 
The outcomes of the latter possibility have been discussed
 in the context of planets orbiting a single evolving star (e.g. \citealp{Livio:1984,Villaver:2007,Nordhaus:2010} and references therein). 

 Although the orbits of both the stellar companion and the 
 planet expand due to mass loss, the change to the planet's orbit is much larger
than the change in the binary's orbit because the mass loss within the planet-star system comprises a greater fraction of the enclosed mass. 
Allowing the planet mass to go to zero, the ratio between
 the semi-major axis of the planet to  stellar companion therefore increases by a factor of:
 \begin{equation}
\left(\frac{a_{p,f}}{a_{*,f}}\right)/\left(\frac{a_{p,i}}{a_{*,i}}\right)=\left(\frac{m_{1,i}}{m_{1,f}}\right)\left(\frac{m_{1,f}+m_{2}}{m_{1,i}+m_{2}}\right).\label{eq:sma-ratio-increase}
\end{equation} 
where the subscripts correspond to the system components ($1,2$, and
$p$ for the evolving star, its stellar companion and its planetary
companion, respectively), and $i$ and $f$ corresponding to the initial and
the final system, respectively.  Relative orbital expansion due to mass loss can shift planets across the stability threshold.

\begin{figure}
\centering
\includegraphics[scale=0.41]{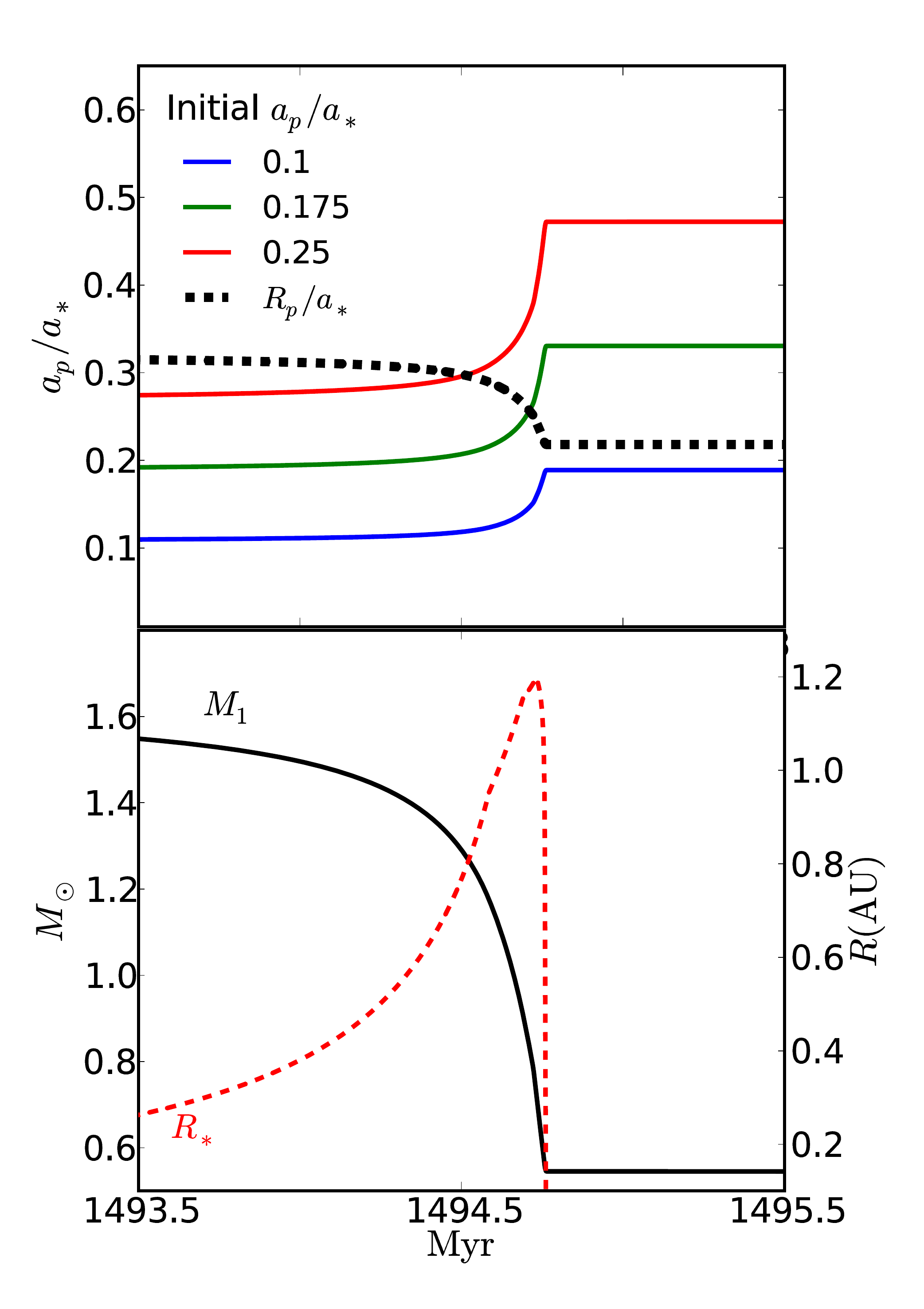}
\caption{  Top panel: Evolution of the ratio of planet to stellar semi-major axis (thin colored lines) and critical stability boundary (thick dashed line) for a mass losing $2\Msun$ with solar mass companion. Initial semi-major axis ratios which cross the stability threshold will enter a chaotic phase of evolution. We show only the  $\sim 2$Myr phase where mass loss is most rapid. The stability threshold declines more sharply than the mass because the stability boundary also shrinks due to the change in $\mu={m_2}/({m_2+m_1})$. Bottom panel: the corresponding stellar mass and stellar radius evolution for the evolving primary.}
\label{fig-destab} 
\end{figure}

 Combining \Eqs{eq:hw-approx}{eq:sma-ratio-increase}, we can define the approximate stability boundary as a function of the evolutionary state (i.e. current primary mass) of a given binary system. In \Fig{fig-destab}, we show the theoretical evolution of several planetary orbits as a function of their separation ratio for the evolution of a $2\Msun$ star in a binary orbit with a $1\Msun$ companion.  In this system, the {\emph{ratio}} of the planetary to stellar orbit, $a_p/a_*$, expands by nearly a factor of two as the primary evolves to become a white dwarf (hereafter WD). At the same time, the critical boundary for stability around the primary shrinks as the primary becomes less massive than its  companion. We show the evolving mass and radius of the primary as well: the stellar model is calculated from the Single Stellar Evolution (SSE) code of \cite{Hurley:2000}, which evolves the primary from $m_1 = 2.0 \rightarrow 0.55\Msun$.

Planets with semi-major axis ratios that exceed the critical point in \Eq{eq:hw-approx} enter into a phase of chaotic evolution. Note that we use chaotic in the technical sense, wherein planets with nearly identical initial conditions undergo dramatically different orbital evolution See \cite{Mudryk:2006} for a discussion of measured Lyapunov exponents for the problem without mass loss.

After becoming unstable, planets can suffer close encounters with either component of the binary. Intriguingly, the instability can also lead to the long-term capture of planets around the secondary star. We now discuss the capture mechanism in more detail.

\section{Capture and the Jacobi Constant}
\label{sec-jacobi-cap}
Before describing our numerical results, we begin with a brief description of the capture process, in which a planet originally orbiting the primary star is destabilized due to stellar mass loss, and then recaptured into a stable orbit around the companion.  We do so by considering the circular restricted three-body problem: coplanar test particles moving about two massive bodies on a circular orbit.

Capture occurs through a process reminiscent of the suggested pull down mechanism for the capture of the Moon \citep{Ruskol:1961} and irregular satellite capture around the gas giants in the Solar System \citep{Heppenheimer:1977,Astakhov:2003}.  In these cases, the body that captures a satellite does so via gaining mass. We show here that mass loss by the primary can also lead to capture.

Capture is controlled by the evolution of a particle's Jacobi constant, $C_J$. Recall that the Jacobi constant is the only conserved integral of motion in the circular restricted three body problem (hereafter CR3BP):
\begin{equation}\label{eq-jacobi}
 C_J = 2\left(\frac{\mu_1}{r_1}+\frac{\mu_2}{r_2}\right)+ n^2(x^2+y^2) - (v_x^2+v_y^2+v_z^2)
 \end{equation}
 where $\mu_1$ and $\mu_2$ are the mass ratios of the stars, $r_1$ and $r_2$ are the separations of the particle from each star, $n$ is the mean motion, and the cartesian components are the positions and velocities of the particle in the non-rotating reference frame \citep{MurrayDermott}. 
 A particle's $C_J$ defines the parameter space in which it may orbit the binary.  The boundaries of the permitted region can be visualized through the use of zero-velocity curves. These contours are obtained by setting the planet velocities to zero and solving \Eq{eq-jacobi} for the positions in the non-rotating reference frame. 
 
 By adding mass loss into the CR3BP, $C_J$ is no longer conserved, because $\mu$ evolves in time. However,  \cite{Lukyanov:2009} has shown that this so-called quasi-Jacobi integral is mathematically well behaved in the case of mass transfer in a binary system, suggesting that it still specifies the bounds of the planet's orbit at any instant.
 
\begin{figure*}
\centering
 \includegraphics[scale=0.59]{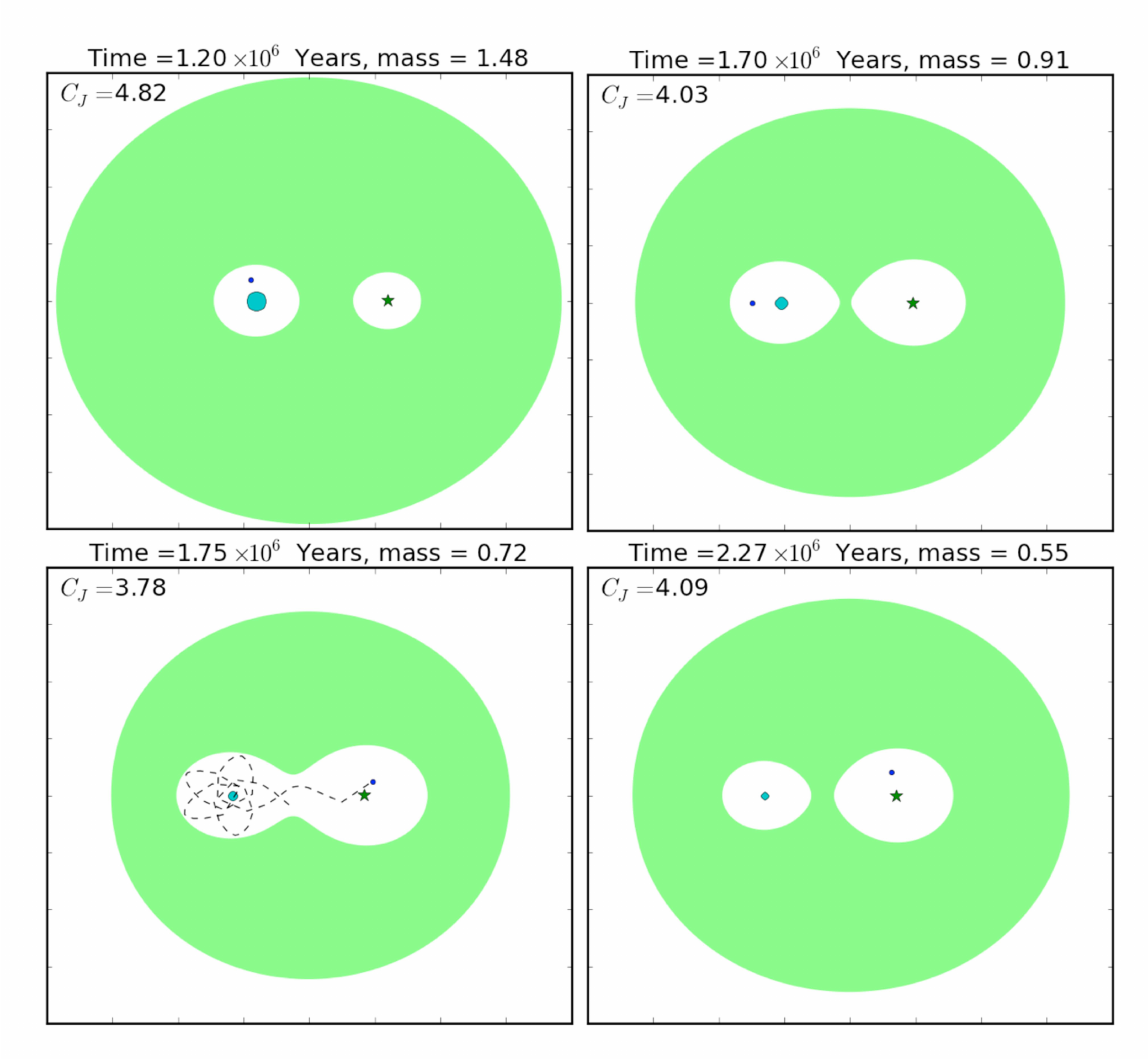}
 \caption{  Evolution of the the zero-velocity curves for a planet as the primary (cyan) loses mass. The white regions indicate where the planet may orbit.  The bottleneck surrounding the first Lagrange point opens midway through the mass loss phase, allowing the planet to orbit both components of the binary. The dashed-tail shows an episode of orbital bouncing, which continues while the curves remain open. By the end of the mass loss phase, the zero-velocity surface has been pinched off, and the planet is trapped around the secondary. A movie of this evolution is available at {\emph{http://www.cfa.harvard.edu/\~{}kkratter/BinaryPlanets/zvelmovie.mpg}}.}
 \label{fig-zvelseq}
 \end{figure*}

\begin{figure}
\centering
\includegraphics[scale=0.46]{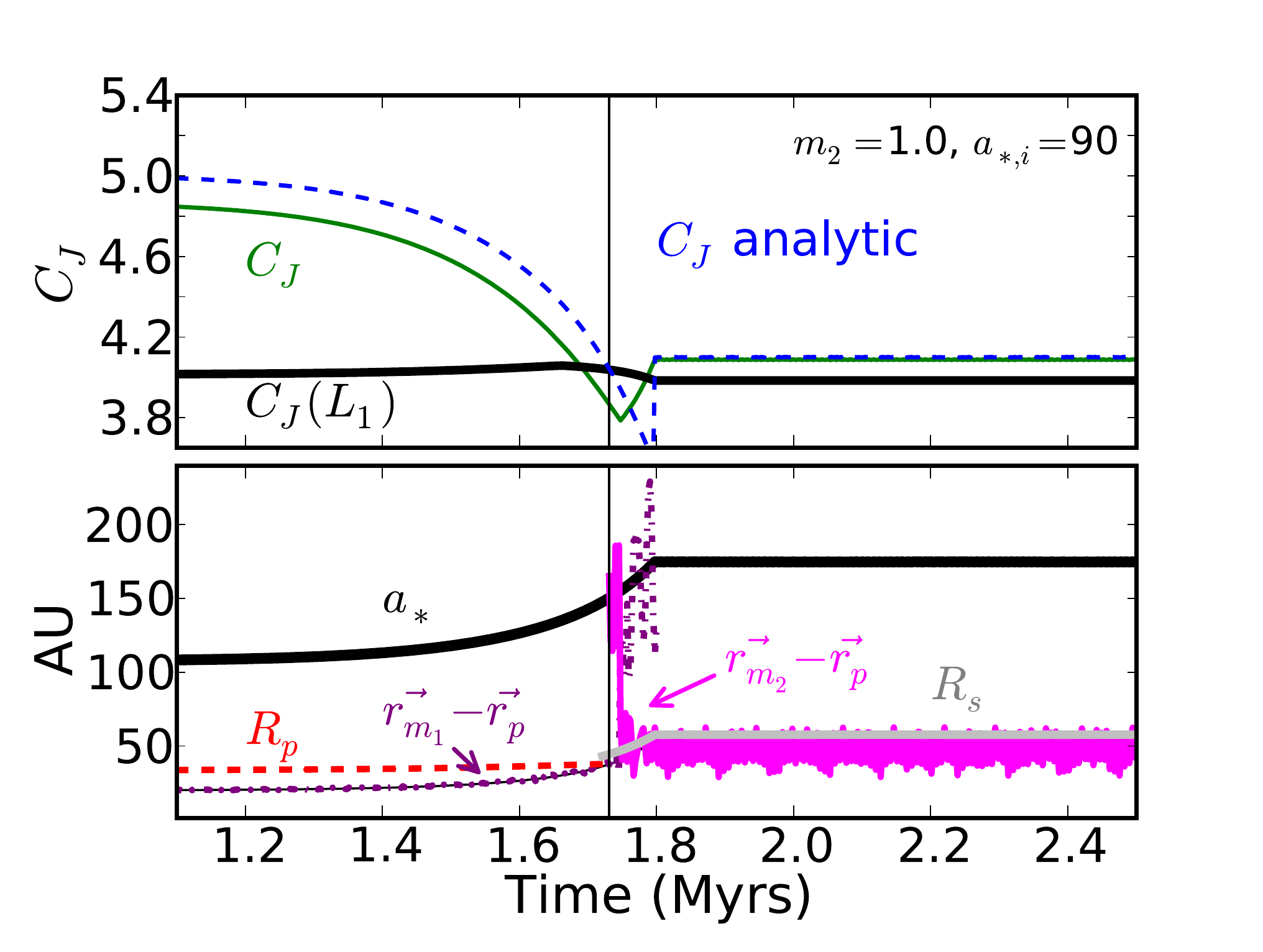}
\caption{Time evolution of the distance and Jacobi constant for the same system shown in \Fig{fig-zvelseq}, where one planet has become trapped in orbit about the secondary. The vertical line shows when the particle crosses the stability threshold defined in \Eq{eq:hw-approx}. The thick black line indicates the stellar separation, the purple line indicates the planet separation from the primary, and the pink line the distance from the secondary. The thick dashed red and grey lines indicate the stability boundaries around the primary and secondary, $R_p$ and $R_s$ respectively. The thin black line shows the initial analytic evolution predicted for the planetary orbit. The top panel shows the particle's $C_J$ (green), the value of $C_J$ at the $L_1$ point (black), and the analytic value for $C_J$ of the particle on a Keplerian orbit (blue dashed), with separation specified by \Eq{eq:adiabatic} as discussed in Section \ref{sec-jacobi-cap}.}
\label{fig-stabex} 

\end{figure}

The evolution of $C_J$ due to mass loss dictates whether a given particle can pass between the stars, or must remain in orbit about one or the other.  \Fig{fig-zvelseq} shows the zero velocity curves for a planet at four epochs in the evolution of a system with $m_1= 2.0 \rightarrow 0.55 \Msun$, $m_2=1.0 \Msun$, and  initial semi-major axis, $a_{*,i} =90$ AU.  \Fig{fig-stabex} is complementary, and shows the evolution of the planetary and stellar separations, as well as the particle's $C_J$ as a function of time. 

The fall and rise of $C_J$ shown in both \Figs{fig-zvelseq}{fig-stabex} is characteristic of host switching. As shown in \Fig{fig-stabex}, we can estimate the evolution of $C_J$ analytically for a particle's orbit that expands with mass loss as dictated by \Eq{eq:adiabatic}. At early times, we keep the planet fixed at its Keplerian rotation velocity. Once the planet nears the instability threshold, the orbit is highly non-Keplerian. In this regime we interpolate between $C_J$ calculated for two assumptions: a) that the planet remains on a Keplerian orbit about the primary, and b) that the planet switches to a Keplerian orbit at the same separation, but about the secondary. The disagreement between the analytic estimate and the actual particle value is worst where orbits depart most from Keplerian (i.e. in the planet hopping region). The relatively better agreement at late times suggests that the particle is well described by a Keplerian orbit, and that the orbital energy is consistent with that of the initial orbit. 

In the case shown here (the simulations are described in the following section),  the planet's $C_J$ at first prohibits transfer between the two stars, then permits it, and then once again forbids it.  If a planet is in the stable region around the secondary when the curves close off, the planet becomes trapped. One can distinguish between open and closed zero-velocity curves by comparing the particles' Jacobi constants to the value at the first Lagrange point ( $L_1$), or:
 \begin{equation}
 C_J(L_1) \approx 3+3^{(4/3)}\mu_2^{(2/3)}-10\mu_2/3
  \end{equation}
 \citep{MurrayDermott}.

Not all systems undergo this complete cycle. For some, when mass loss stops, the zero velocity curves remain open, and capture is less likely. Note that the stability threshold defined in \Eq{eq:hw-approx} is crossed only after the curves have opened: planets can remain on stable orbits in the face of open zero-velocity curves. 

  We show in Section \ref{sec-results} that the relative value of the the average Jacobi constant of the surviving particles and at the $L_1$ point, $\bar{C_J}-C_J(L_1)$, correlates with the probability of planets being captured. 

\section{Few-body simulations}
\label{sec-nbody}

We use few-body simulations to explore the evolution of planets in evolving binary systems.  We focus on a limited phase space, which allows us to explore the complex dynamical behavior as a function of system parameters and mass loss rates.  We consider only point masses, and limit ourselves to circular binaries hosting initially circular, coplanar planets, and a single primary mass and evolutionary pathway. Initial planet eccentricity is unlikely to be important, as shown by \cite{Mudryk:2006}. 

 We use a fourth order Hermite integrator based on the algorithm of  \cite{Hut:1995}, with an additional mass loss term.  Mass loss rates and evolving stellar radii are calculated from the SSE models of \cite{Hurley:2000}. We use the default values for mass loss rates (assuming solar metallicity) provided in the publicly available version of the code. Each run consists of two massive bodies, and 100 test particles representing planets. The time step for all particles is set to $10^{-2}-10^{-3}$ of the encounter timescale for the closest pair of particles in the simulation, where the smaller step is required for good energy conservation for the fastest mass loss rates\footnote{ The outcomes remain the same under an order of magnitude reduction in the time step.}. The stellar mass is updated on the same time step as the test particles, and all particles share the same time step. When mass loss is included, energy is not conserved by default, however in test runs with no mass loss, the code conserves energy to 1 part in $10^{12}$ over a 5 Myr run. We also confirm that particles on both circular and chaotic orbits (e.g. bouncing between the stars) conserve their Jacobi constants over a 5 Myr run (see \citealt{Moeckel:2012}). We check that the evolution of the stars and test particles follows the expected orbital expansion ( Equation \ref{eq:adiabatic}) up until the point of instability.  Note that for the realistic (fast) mass loss case, a small eccentricity is introduced into the stellar binary -- this results in oscillations in the Jacobi constant for some surviving test particles.

\subsection{Simulation Details}
We simulate evolving binaries containing primaries which evolve from  $M=2\rightarrow 0.55\Msun$. Such A stars host planets \citep{Johnson:2007}, undergo extensive mass loss in a Hubble time and are frequently found in multiple systems \citep{Raghavan2010}. 
We consider secondary masses ranging from $m_2=0.5-1.7\Msun$, and initial binary separations from $a_{*,i}  =75-105$ AU. We fix the planet separation at 15 AU: far enough to avoid any interaction with the primary's envelope, and close enough to be well inside the (initial) stability region around the host star. Test particles are initialized with zero eccentricity, random phase, on coplanar orbits.  In order to expedite the calculations, we evolve the planetary and stellar orbits with the analytic formulae for the first $\sim 1490$ Myr, until the star reaches the beginning of the asymptotic giant branch (AGB), at which point the mass loss rate rapidly increases (shown in \Fig{fig-destab}).

We consider only the evolution of the more massive star, which leaves the main sequence first (the planet host). In Section \ref{sec-results} we discuss possible outcomes when the second star evolves off of the main sequence.

 In total we explore 70 different binary configurations. For each binary configuration, we also explore the influence of mass loss rate and collisional cross section. Even varying only two parameters produces a complex set of outcomes because the probabilistic fate of a planet depends not only on the initial and final configurations, but also on the rate at which they pass through each state, which is determined by the mass loss rate and expansion of the primary's atmosphere.   
We run our fiducial calculations for 30 Myr following the end of the mass loss phase, or $10^4$ orbits, as most planets that survive to 10 Myr remain out to 100 Myr in several test cases. As seen by \cite{Rabl:1988} and \cite{Holman:1999}, in the case with no mass loss, most of the orbital evolution occurs on timescales of less than $300$ binary orbits. We discuss these timescales further in Section \ref{sec-collisions}.

Even though all the particles begin with nearly identical $C_J$ (small variations are introduced due to variations in the planets' initial phases), mass loss induces a spread in $C_J$ because they undergo different chaotic evolutionary paths while mass loss is ongoing. This spread produces a range of outcomes for planets in a given system.

During the course of the simulation we keep track of each particles closest approach to either star. Particles which penetrate the specified stellar radii are removed from the calculation and identified as collisions. For the primary, the collisional radius  is set to be the maximum of either the stellar radius calculated from \cite{Hurley:2000} or the star's approximate tidal radius  as a WD:  $3.8\times10^{-3}$ AU. The secondary's radius is fixed at $1\Rsun$ for all masses to simplify the interpretation of probabilities. 

Because we neglect tidal effects, we also identify any particle that has suffered a close approach within 3 tidal radii of either star (but not a direct collision) as having had a tidal interaction. We define the tidal radius to be:
\begin{equation}
r_{t}\approx\left(\frac{M_{\star}}{M_{p}}\right)^{1/3}r_{p}.
\end{equation}
Particles that do not suffer tidal interactions or collisions are identified as bound to either the primary or secondary, although typically no planets remain around the primary\footnote{As shown in Table \ref{tab-res}, there are five systems which retain planets around the primary: in one case ($m_2=0.8 \Msun, a_{*,i} = 105\rm{AU})$, the system evolves up to the edge of the stability threshold, and because of phase variations only some planets become unstable. We repeated this run several times and found similar results. For the remaining systems, those classified as bound to the primary are on bouncing orbits that we expect will ultimately result in collisions: they happen to be in orbit about the primary at the conclusion of the simulation.}.
\begin{figure*}

\centering
\includegraphics[scale=0.55]{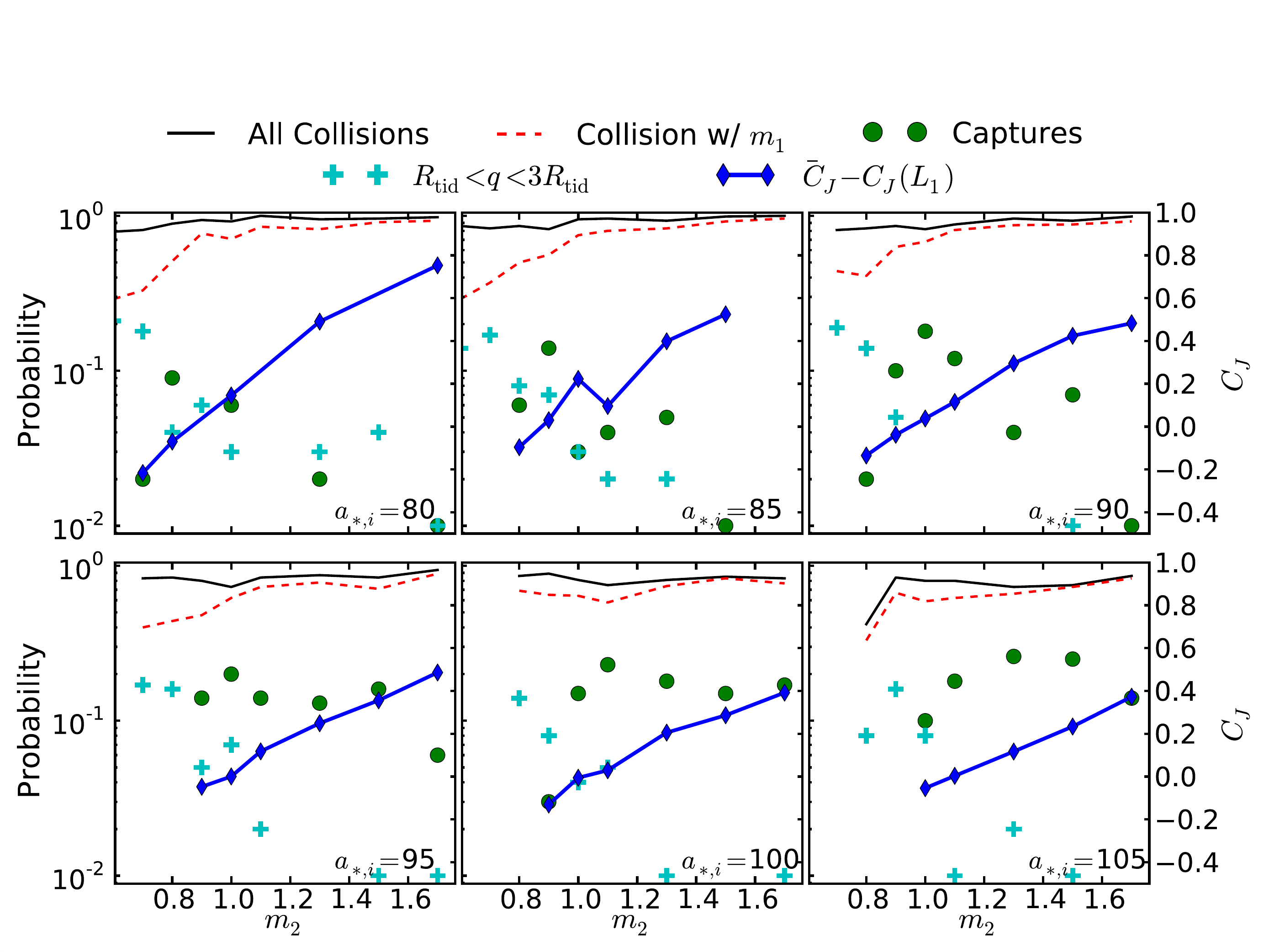}
\caption{   Probabilities for collision, capture, and tidal encounters in our fiducial models. Masses are in units of $\Msun$ and separations in AU. Each panel represents a different semi-major axis for the binary, while different secondary masses are shown along the x-axis of each panel. All collisions (black), and collisions with the primary (red dashed) typically increase with increasing mass of the secondary, and decrease with increasing binary separation. The decrease in tidal encounters (cyan plusses) with increasing secondary mass is due to more collisions with the AGB atmosphere due to the earlier onset of chaotic orbits. Whether or not the zero-velocity curves have pinched off, preventing continued planet bouncing, is indicated by the sign of ${\bar{C_J}-C_J(L_1)}$ (blue diamonds). Positive values show that the curves are closed.  }
\label{fig-6panel-real}
\end{figure*}

\section{Results}
\label{sec-results}

We begin by showing the probabilities of capture, collision, and tidal encounters for the fiducial runs with realistic mass loss rates and stellar radii in \Fig{fig-6panel-real}. We then describe in turn the importance of the mass loss rate, and physical radius in controlling captures and collisions.

\subsection{Outcomes from fiducial few-body simulations}\label{sec:fboutcomes}
\Fig{fig-6panel-real} shows the probabilities for captures, collisions, and tidal interactions for 60 of our 70 binary systems. The tabulated results for all systems are shown in Table \ref{tab-res} in the Appendix.  In the figure, each panel shows the outcome for a fixed initial binary separation for a range of masses. For all systems that become unstable, the most likely outcome for a planet is collision with one of the stars. 

Several trends are readily observed. Collisions typically increase towards higher masses at fixed separation, while tidal encounters decrease. Both trends are the result of the instability threshold being reached at an earlier evolutionary state, when the primary has an inflated radius;  wider encounters become collisions, and encounters within a few stellar radii are prohibited. Note that in order to collide with either star, planets must be set onto low angular momentum orbits. As a result, collisions involve at least one passage back and forth between the stars.

At larger separations and stellar masses, planets are frequently ($10\%$) captured by the secondary star. Captured particles come in two varieties: those whose energies as measured via $C_J$ dictate that they remain in orbit only about the secondary, and those who in principal may pass back into orbit around the primary, but show no evidence of such movement over the course of the simulation. These two are distinguished in \Fig{fig-6panel-real} by the sign of $\bar{C_J}-C_J(L_1)$, where only those with positive values are truly trapped (closed zero velocity curves). Clearly, some with open curves (negative values) do survive for the duration of the simulation. This result is in line with the findings of \cite{Heppenheimer:1977}. They attribute this behavior to planets entering into special orbits that avoid the conjunction of apocenter with the $L_1$ point. In general, more planets are captured in systems whose zero-velocity curves close at the end of mass loss.

Both classes of captured planets typically have prograde, eccentric orbits ($e> 0.4$), and rapid, retrograde precession in the frame of the binary. Using the method of the surface of section, we identify many of these planets as having quasi-periodic, though formally chaotic orbital parameters \citep{MurrayDermott}. Their long term ($>100$Myr) survival remains uncertain. Nevertheless, weaker tidal effects may also dissipate energy, causing such orbits to shrink and circularize \citep{Moeckel:2012}. 

 The captured planets typically have orbits which graze the stability boundary,  as shown in \Fig{fig-stabex}. Such planets are extremely likely to go unstable when the second star in the system evolves to become a WD. A secondary instability cycle might re-populate the original host with planets on eccentric orbits. 
 
Close approaches, $q$, within $1 R_* < q  <3 R_*$, are also common (~10\%), we discuss our treatment of these planets, and the plausible results of close encounters and physical collisions below.

\subsection{Tidal encounters}
\label{sec-tidal}
\begin{figure}
\centering
\includegraphics[scale=0.43]{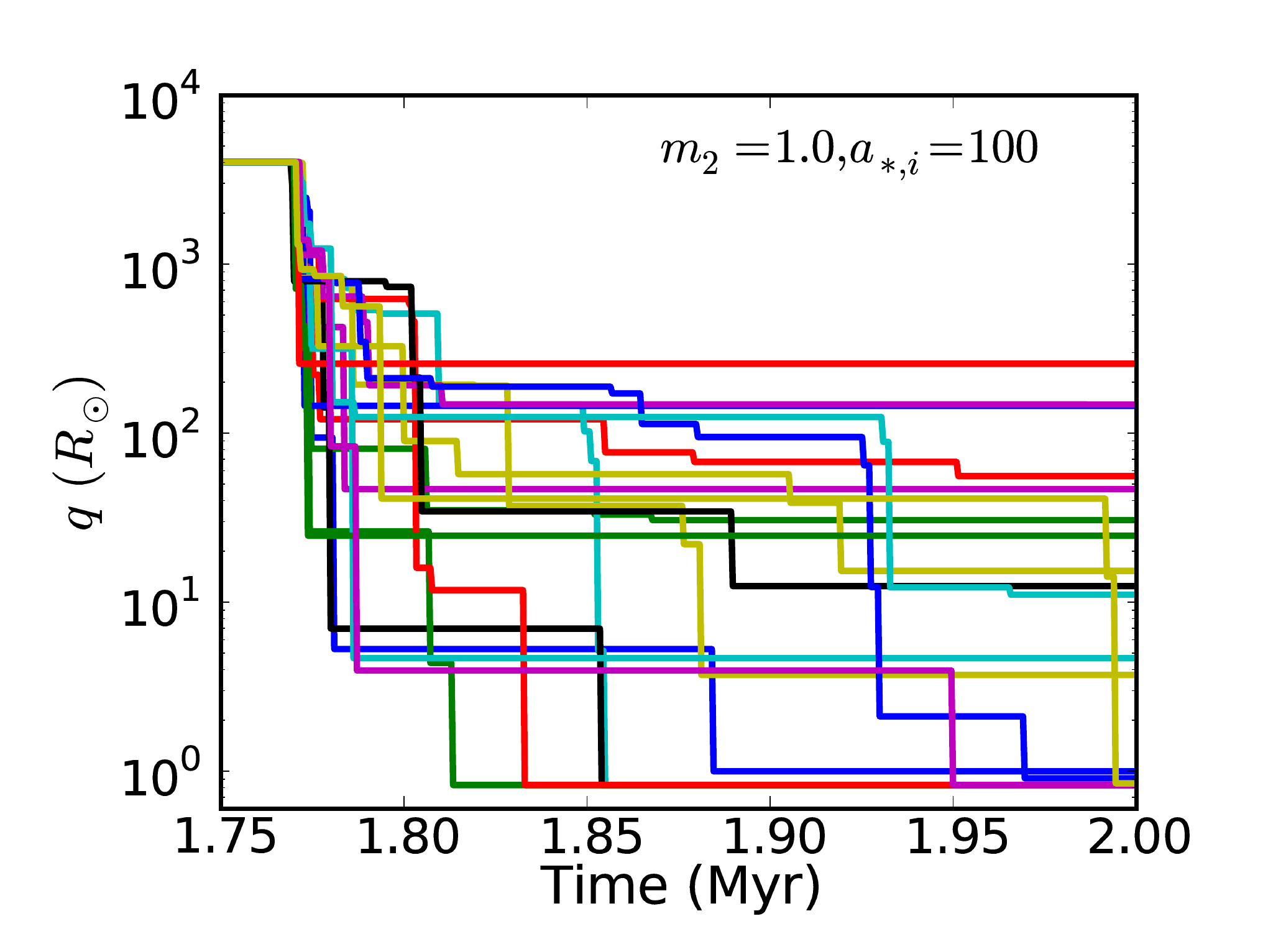}
\caption{  Closest approach $q$ in units of solar radii for a sample of test particles in one binary system. Each color represents a different particle trajectory. Note that particles typically undergo large jumps in close approach, suggesting that their orbits are chaotically sampling the orbital parameter space, not decaying to ever smaller separations.}
\label{fig-jumprad}
\end{figure}

 By excluding planets with previous close approaches ($ q\ltsim3R_\odot$) from our calculated probabilities for both capture and collision, we expect that our neglect of tidal effects is unlikely to bias our results. When particle orbits are chaotic, they do not evolve smoothly to ever closer values of $q$, but rather make order of magnitude jumps in separation from one close approach to the next.  To demonstrate this behavior, we show in \Fig{fig-jumprad} the evolution of the closest approach distance for a small sample of particles from one run. We see that a particle that reaches a separation of a few stellar radii has likely not had previous encounters at comparable distances, and thus not experienced significant tidal damping. For planets that suffer encounters within a few stellar radii, tidal disruption is the most likely outcome.

\subsubsection{Tidal disruption versus capture}
Planet encounters at many stellar radii lose almost no energy because the energy dissipated scales as $a_p^{-6}$, where $a_p$ is the periastron. When tidal effects become important at a few tidal radii, close approaches sap energy from the planet's orbit. In principal these encounters can decrease the planet's velocity (and thus increase $C_J$),  enough to bring it from an unstable to stable orbit in a binary (or from unbound to bound in a single star system). Tidal capture has been suggested as a formation mechanism for X-ray binaries through close encounters between stars and compact objects in globular clusters \citep{fab+75},  and for close-in planets \citep{Faber:2005}. 

Recent hydrodynamical simulations suggest that close approaches lead to the disruption or ejection of a Jupiter-mass planet, rather than long-term survival \citep{gui+11}.  At relatively larger distances,  energy is dissipated through tidal heating of the planet, causing its orbit to shrink. As the orbit decays, the planet begins losing mass through tidal stripping; this stripping and mass transfer typically lead to the ejection of the planet to larger separations at high velocity, or to total planetary disruption.

The binary scenario is somewhat different than the case of a single-star planet interaction. Tidally stripped planets that would have been ejected in the single-star scenario could still be bound to the binary system, and therefore continue to evolve chaotically following their ``ejection'' in the tidal encounter. In this case, the planet might survive or suffer a collision just as any other planet in the system. These ``shaved", heated planets might be much brighter than their cooling age.
A direct physical collision with a MS star, (in our case, the secondary) should also lead to destruction, as the tidal radius of Sun-like stars is comparable to their physical radius for Jupiter-like planets. Physical collisions might also be directly observable with upcoming transient surveys \citep{Metzger:2012}.

In the case of an encounter with the primary as a WD, the tidal radius is much larger than the physical radius of the WD; thus disruption is likely prior to a direct physical collision. Disruption can lead to the formation of an accretion disk around the compact object, and the infall of material onto it. Such debris disks around WDs potentially serve as a source for long lasting metal pollution (e.g. \citealp{kle+11} and references therein).  Similar encounters have been studied in the context of the tidal disruption of a star by a massive black hole (e.g. \citealp{ree+88}). 

\subsection{Collisions with AGB Atmospheres}
\label{sec-collisions}
\begin{figure*}
\centering
\includegraphics[scale=0.46]{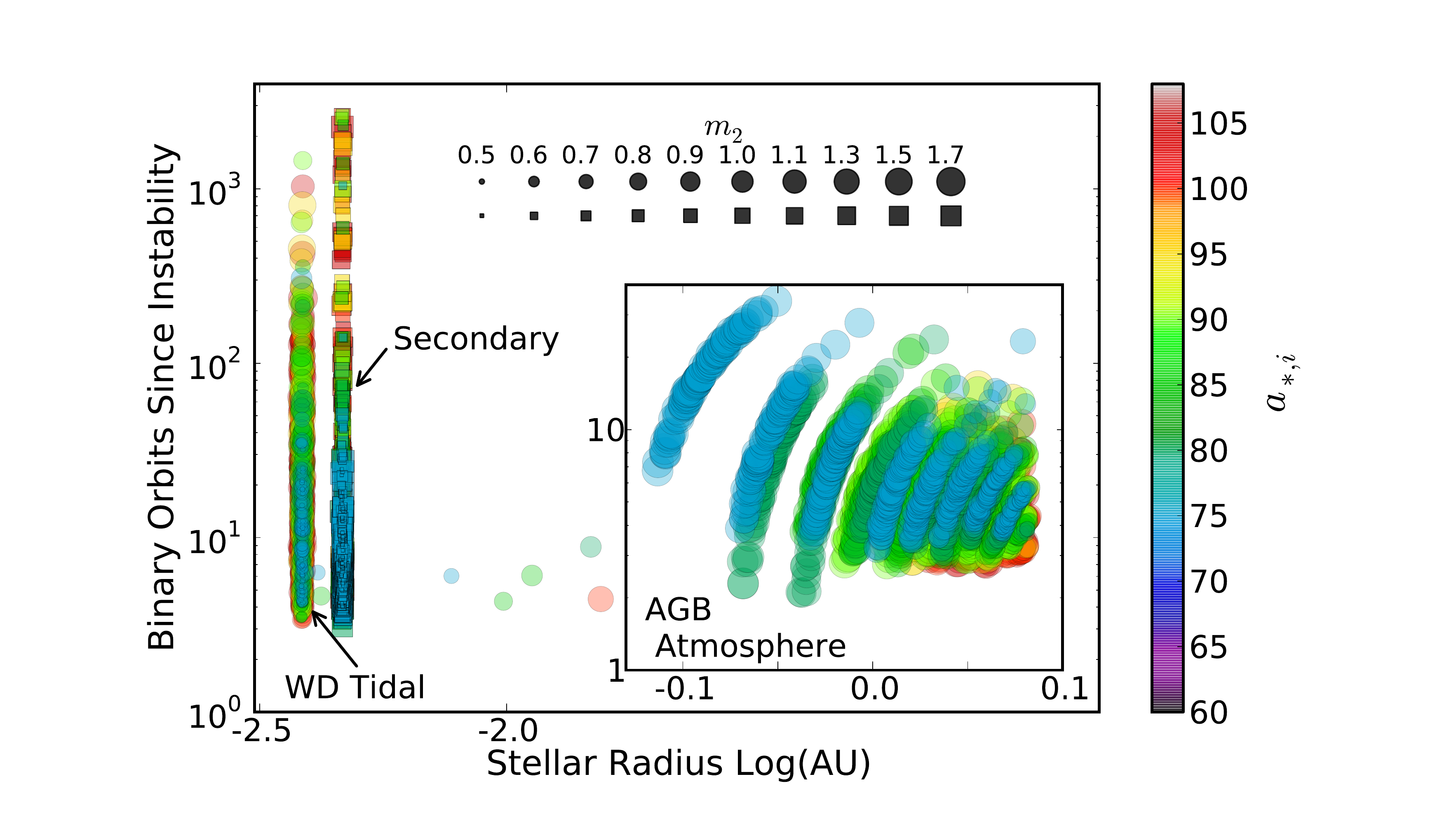}
\caption{  \label{fig-collradii} Distribution of collision times and radii for the fiducial set of runs with realistic radii and mass loss rates. The inset panel is zoomed in on the radius range of the AGB atmosphere. Masses are in unites of $\Msun$, and separations in AU. Color indicates the initial binary separation, and symbol size the mass of the secondary. Time is measured in units of each host binary's orbital period, with $t=0$ set to be the point at which the system violates the instability criterion in \Eq{eq:hw-approx}. Collisions with the primary are indicated by circles, while those with the secondary are shown with squares. A movie which shows the population of this phase space is available at {\emph{http://www.cfa.harvard.edu/\~{}kkratter/BinaryPlanets/radcollmovie.mpg}}.}

\end{figure*}

AGB stars spend a short time ($\sim10^5$ yrs) inflated to radii of order 1 AU. Consequently we find that the radius distribution of collisions with the primary is highly bimodal: they either occur at radii near the maximum, or the minimum radius in our models--the tidal radius of the WD (see \Fig{fig-destab}). \Fig{fig-collradii} shows the distribution of collision radii and times for all collisions in our fiducial models. We measure time in units of the host binary's orbital period at the onset of instability, and $t=0$ is set to the moment when the system crosses the instability threshold.  We can identify several trends. First, the bimodality of the distribution is easily observed: only 0.1\% of collisions occur at intermediate radii. Secondly, we can see that closer binaries are destabilized earlier, and therefore collide with the AGB atmosphere at slightly smaller sizes while the atmosphere is still expanding (note the blue points are shifted to smaller radii compared to the green and red points). Lower mass secondaries produce AGB collisions only at close separations, whereas more massive secondaries drive the system unstable quickly enough to drive AGB encounters at all binary separations considered. We see that most collisions occur within the first  $\sim 200$ binary orbits, as seen by \cite{Holman:1999} and \cite{David:2003}. We discuss collision timescales in more detail in the Section \ref{sec-longtimes}.

There are several plausible physical outcomes following collisions between the planet and AGB star.
Collisions with a partially evolved star might lead to the engulfment of the planet in the stellar envelope, and ultimately to its in-spiral and disruption inside the star. In the process, the planet might transfer energy, angular momentum, and heavy elements into the stellar envelope.

Collisions with the more loosely bound AGB atmosphere likely lead to a qualitatively different outcome. Such interactions might form a low mass cataclysmic binary system as the planet penetrates the AGB envelope before its ejection and accretes from it \citep{Livio:1984}. It can also affect the structure of the planetary nebulae that ultimately forms around the WD remnant \citep{sok01}, in a similar manner to stellar binary companions.  
The planet might also completely eject the envelope, depending on the its penetration depth, somewhat similar to the case of a common envelope scenario. In this case,  kinetic energy dissipated in the collision and envelope ejection could be sufficient to capture the surviving planet into a close orbit around the stellar remnant. This latter scenario provides a unique possibility for placing planets in tight orbits around white dwarfs (see also \citealp{bea+11} for a related discussion).  Note that if tidal dissipation leads to orbits near the tidal radius of the WD, such systems would reside in the WD habitable zone \citep{Agol:2011}.

\subsection{Varying mass loss rates}
\label{sec-longtimes}
Mass loss due to stellar winds occurs on a timescale that is slow (adiabatic) compared to the planet and stellar orbital period.  However, there is another relevant timescale in the calculation: the average time for a planet to reach its final ``outcome" after becoming unstable,  i.e. collision or a quasi-stable orbit. If the stellar mass and binary separation evolve on this timescale, then the probability for a collision or capture is determined by the integrated probabilities at each state over the time spent at each state.

In order to isolate the effect of system mass ratio and separation, and thus determine probabilities as a function of system parameters exclusively, we run two sets of comparison simulations with unrealistically long mass loss timescales set to a constant mass loss rate over  2 Myr, and 20 Myr. We fix the collision radius to the solar radius at all times.  In both cases, we compute the timescale  over which  $75\%$ of particles destined to collide suffer a collision; this serves as our ``equilibrium timescale,'' and ranges from  $ 0.2 \ltsim t_{\rm coll} \ltsim 1$ Myr, depending on the system. Changing the fraction of collisions which defines our equilibrium timescale by 10\% has minimal effect on the timescales. Thus only the 20 Myr mass loss timescale case evolves adiabatically with respect to all timescales in the problem. 

In this limit, nearly all (99\%) of particles suffer a collision in all of the systems that become unstable. That collisions increase in the long mass loss timescale limit is easily seen from the evolution of the planets' $C_J$. At the onset of instability, the $L_1$ bottleneck is near its peak width. This facilitates exchanges between stars, which in turn are responsible for setting planets on collisional orbits. Capture occurs as the bottleneck closes off, which only happens at the tail end of mass loss. In the case where the mass loss timescale is long compared to the collisional time, all of the particles will undergo sufficient exchanges to collide prior to the closing of the zero-velocity curves, and therefore few are left at the end of the evolution.  We now use these long timescale runs to show that the relative collision probability between the primary and secondary scales inversely with mass ratio.

\subsection{Collision Rates: Long mass loss times} 
\begin{figure}[htbp]
\begin{center}
\includegraphics[scale=0.43]{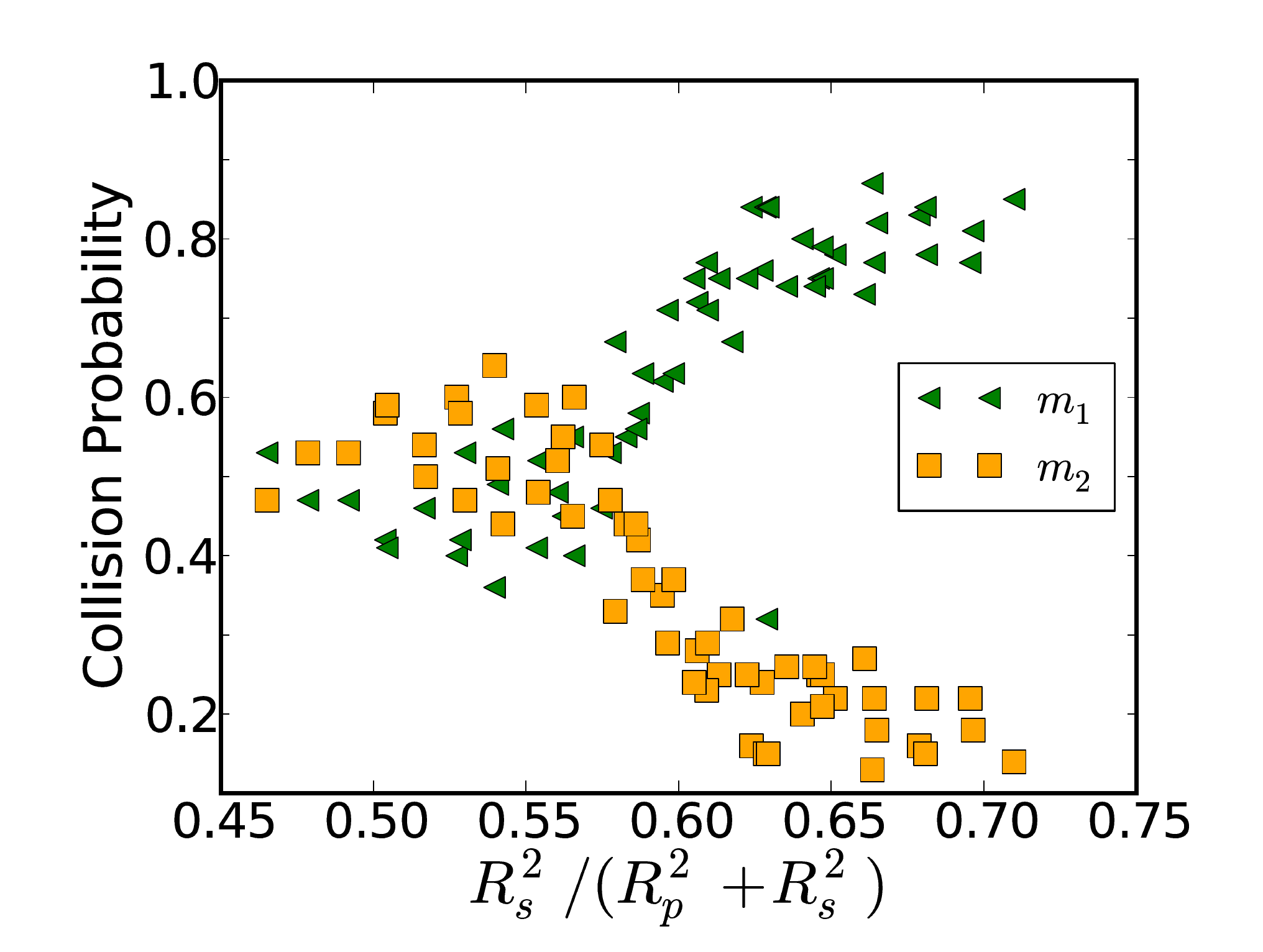}
\caption{ \label{fig-coll_prob} Collisional probability for $m_1$ and $m_2$ as a function of the relative area of the secondary stability region $R_s$ to the total available stable area: $R_s^2/(R_p^2+R_s^2)$. These probabilities are calculated for the simulation set with slow mass loss (20 Myrs), and constant radius. The lower bound on area for which we show collisions illustrates the minimum relative areas achieved for comparable primary and secondary masses.}
\end{center}

\end{figure}

Once planets become unstable, in the process of bouncing back and forth, they sample the orbital parameter space about each component, defined by the area enclosed within the zero velocity curves.  We find that the relative collision rates between the stars are well correlated with the area of the stable region around each star as defined by the critical radius in \Eq{eq:hw-approx}. 
 When the stability radius $R_s$ or $R_p$ is smaller, the physical stellar radius subtends a larger fraction of the allowed orbital area, and thus there is a higher probability for suffering a collision.  The area ratio depends on the square of the stellar mass ratios. Collision probabilities scale inversely with the area around the star as shown in \Fig{fig-coll_prob}, meaning that collisions occur predominantly with the star that is less massive. 

\begin{figure}[htbp]
\begin{center}
\includegraphics[scale=0.43]{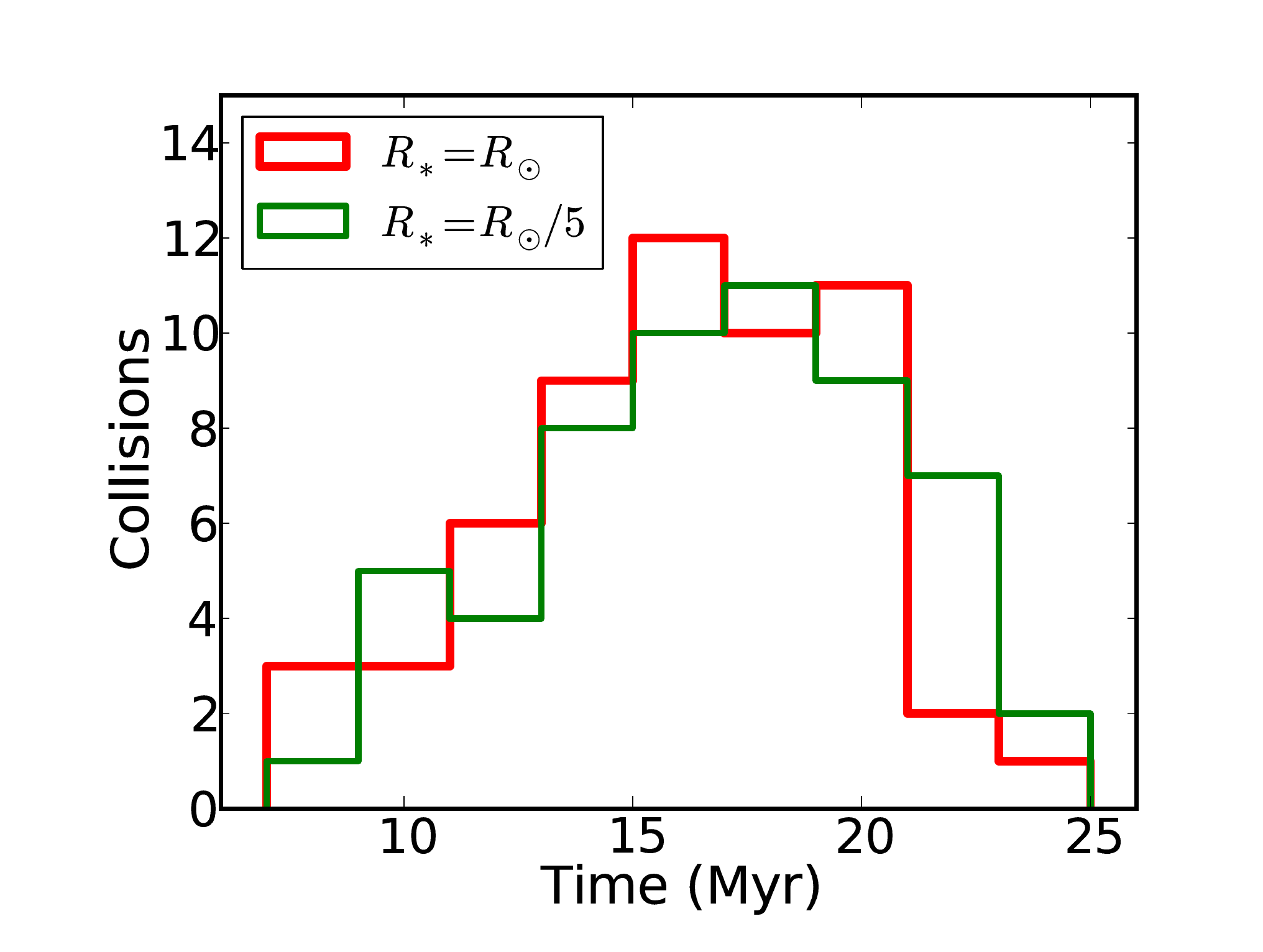}
\caption{\label{fig-coll-time} Distribution of collision times for the slow mass loss (20 Myrs) runs, with collision radius set to  $R_{\odot}$ (red), and $R_{\odot}/5.$ (green). As expected, the collision rate peaks slightly earlier for larger stellar cross section. Nevertheless, the relative similarity demonstrates that colliding particles are on sufficiently low angular momentum orbits that collisions destined to reach a stellar radius will penetrate to 1/5th of that distance. }
\end{center}

\end{figure}

In these runs, the stellar radii are fixed at $1\Rsun$ throughout the calculation. We also run the same models with the collision radius reduced by a factor of 5, and find that the collision probabilities remain at $99\%$ across all unstable systems, although the collision times are shifted to slightly larger values, as shown in \Fig{fig-coll-time}. Our interpretation of this trend is that particles set on collisional trajectories have nearly radial orbits. This property, and the corresponding high collision rate might well  be the result of considering only coplanar planets. 

\subsection{Free floating planets: ejection}
In the results presented here, destabilized planets either collided with one of the binary components, or exchanged hosts. No planets were ejected from the system. This is because the minimum $C_J$ reached in the systems always prohibited escape. For lower $C_J$, zero-velocity curves take on a ``c" shape, opening up to the circumbinary orbital region, thus allowing escape. In several exploratory integrations of non-coplanar systems ($i \sim 10^\circ$) and systems with more unequal initial binary mass ratios ($\mu \sim 0.1$),  we found that planetary ejections are possible.  For lower mass secondaries, lower $C_J$ may be reached because the secondary and planet can be more tightly packed prior to instability. Non-coplanar orbits likely allow for lower $C_J$ through an expanded phase space in which eccentricity (and inclination) excitation increase the velocity of the test particle at a fixed location.  We defer further discussion of escaping planets to future work.    

\section{Frequency of Destabilized Planets}
\label{sec-frequency}
We have thus far shown that the planetary TEDI can produce a wide range of astrophysically interesting configurations. We now try to estimate the occurrence rate of planet destabilization. At present, there are few constraints on both the overall frequency of planetary systems in binaries,  and on wide orbit systems in general. Nevertheless, we can calculate the frequency of destabilized systems for various planetary orbital distributions. Given the simplified approach we take, the results should be interpreted with caution; they are dependent on highly uncertain model assumptions.

Following \cite{PK2012}
we construct a population of stellar binaries, whose properties are chosen randomly from the appropriate 
observed distributions for A-G stars \citep{Raghavan2010}.
The periods are chosen from a distribution of orbital
separations that is log-normal. The mass ratio of the binary, $m_{1}/m_{2}$, is chosen from
a Gaussian distribution (mean of $0.6$ and dispersion of $0.1$).
Following \cite{fab+07},
the distribution of initial eccentricities is a function of period \citep{duq+91}. For periods
shorter than 1000 days, the eccentricity was chosen from a Rayleigh
distribution {[}$dp\propto e\exp(-\beta e^{2})de${]} with $\left\langle e^{2}\right\rangle ^{1/2}=\beta^{-1/2}=0.33$. For periods longer than 1000 days, the eccentricity was chosen from
an Ambartsumian distribution ($dp=2ede$), corresponding to a uniform distribution on the energy surface in phase space.

We populate the binary systems with single planets. Since the distribution of planet orbital properties is unknown, we consider three cases : (1) period distributions similar to that of low mass binaries (log-normal), (2) period distributions similar to that of high mass binaries (log-uniform), and (3) uniform distribution of periods. The stability criteria derived by \cite{Holman:1999} does not account for the planetary eccentricity, and we therefore do not specify it. 

For the purposes of calculating frequencies, we integrate our planet distribution down to  $3R_\odot$ from the primary star,  however we consider all planets at initial separations below 7 AU to be ``stable'', in that they may interact with the stellar envelope or wind, and thus will not follow the simple adiabatic mass loss evolution envisioned here \citep{Villaver:2009}.

For all wider systems, we use Equation  (1) in \cite{Holman:1999} (which includes the binary eccentricity dependence) to verify whether the initial planetary system is stable at birth, and if it is not, we generate a replacement system. We construct 1000 initially stable planetary systems for each primary mass bin ranging from $1-3\Msun$ for each of the three planetary orbital distributions. Each primary mass bin is $0.1$ M$_\odot$. 

We then use the initial to final mass relation determined by \cite{Salaris:2009}, to find the final mass of the primary star as it becomes a WD. Using \Eq{eq:adiabatic} to calculate the evolved orbits of the binary and planetary companions, we check whether the planetary system has crossed the instability threshold.

\begin{figure}[htbp]
\begin{center}
\includegraphics[scale=0.43]{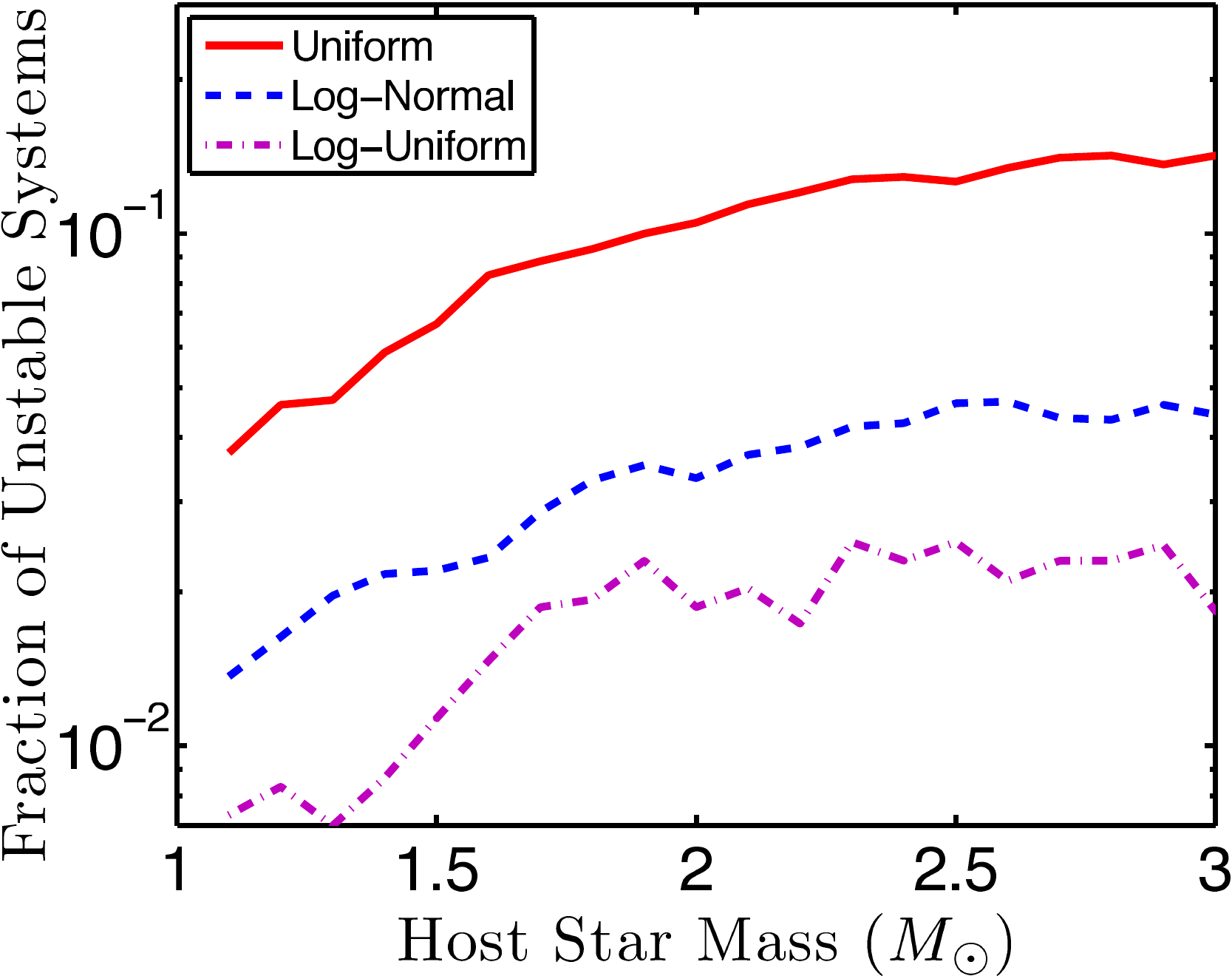}
\caption{\label{fig:unstable_fracs} The frequency of destabilized planetary systems as a function of the primary host mass. Three cases are shown, corresponding to a log-normal, log-uniform and uniform distribution of planetary orbital periods.}
\end{center}

\end{figure}

The total fraction of unstable systems is recorded; the results are shown in Fig. \ref{fig:unstable_fracs}. This figure provides an estimate for the fraction of destabilized planetary systems for a given primary mass host. Only primaries more massive than 1$\Msun$ become WDs in a Hubble time, and we therefore use this as a lower limit. Overall we find that a few up to $\sim10\%$ of all WD binary systems that contained planets at birth might undergo the TEDI.

\section{Summary} 
\label{sec-discussion}
In this paper we have examined the effect of binary stellar evolution on the fate of wide-orbit planets.  We have shown that a substantial fraction of binary planetary systems may be susceptible to the planetary version of the Triple Evolution Dynamical Instability \citep{PK2012}, where mass loss driven orbital expansion causes planetary orbits to become unstable and chaotic. We have considered circular binary systems and coplanar planets so that we may exploit the simplicity provided by the circular restricted three-body problem.

As a result of the chaotic orbits induced by mass loss, planets can suffer close encounters or collide with either component in the binary. When close encounters occur while the mass losing star is on the asymptotic giant branch, dissipation of energy in the envelope can potentially lead to planetary captures on close orbits. Such planets might reside near the tidal radius of the white dwarf following envelope ejection; these planets would fall within the white dwarf habitable zone, and are potentially discoverable with future transit surveys \citep{Agol:2011}. 
We note that tidal heating might provide a source of geothermal energy to such planets. Recent numerical work by \cite{Kim:2012} also finds that the mere proximity of a planet to the low density atmosphere might excite spiral arm structure observable with ALMA. Encounters within a few tidal radii of either the main-sequence secondary, or the newborn white dwarf likely result in planetary disruption. Planetary disruption around the white dwarf might contribute to the observed atmospheric pollution \citep{Farihi:2009,Dufour:2010,kle+11,{Bochkarev:2011}}. 

Planets that do not suffer close stellar encounters can become captured by the secondary. Mass loss and planet hopping drive a fall and then rise in the planets' Jacobi constants. Mass loss first allows planets to hop back and forth between the stars; in some cases the late time rise in the Jacobi constant causes planets to become captured in orbit about the secondary on quasi-stable orbits (see \Fig{fig-zvelseq}). Because such planets reside near the boundary of instability, when the lower mass star in the binary evolves, a second round of instability is possible.

We suggest that future surveys for planets in evolved binary systems might find:
\begin{itemize}
\item {planets around the main-sequence star in a WD-main-sequence binary at the edge of orbital stability}
\item{planets around the WD in a WD-main-sequence binary on sufficiently close orbits that they would have been engulfed in AGB atmospheres in standard evolutionary models}
\item{planets around the more evolved WD in a WD-WD binary at the edge of orbital stability.}
\end{itemize}

Future work must constrain in detail the outcomes of collisions between planets and AGB atmospheres to make more detailed predictions for the types of orbits produced by this interaction. Similarly, an exploration of multi-planet, and non-coplanar systems is required to provide more robust observational predictions.

\acknowledgments
The authors would like to thank Dimitri Veras and Andrew Youdin for comments on an earlier draft of this manuscript, and Nick Moeckel, Matt Holman, Smadar Naoz, Fred Rasio, and Greg Laughlin for helpful discussions. The authors thank the referee for helpful comments. KMK is supported by an Institute for Theory and Computation fellowship through the Harvard College Observatory. HBP is a BIKURA (FIRST) and CfA prize fellow.

\appendix

We include in tabular form the outcomes for all binaries in our fiducial models. We classify particles according to three possible outcomes: bound, colliding, and tidally interacting with one of the two components. Note that in several cases, the total number of interactions sums to greater than the number of particles in the simulation (100). This is due to the fact that several planets suffer tidal encounters with both components. We discuss these results in Section \ref{sec:fboutcomes}.

\begin{deluxetable}{c c c c c c c c}

\label{tab-res}
\tablecaption{Tabulated results from the fiducial suite of integrations with 100 test particles in each binary. Particles are classified as either bound, colliding, or tidally interacting with either the primary or secondary.}

\tablehead{Mass ($\Msun$) & Sep. (AU) & Bound 1 & Bound 2 & Coll 1 & Coll 2 & Tid 1& Tid 2}
\startdata
0.5&75&0.0&0.0&34.0&50.0&11.0&5.0\\
0.5&80&100.0&0.0&0.0&0.0&0.0&0.0\\
0.5&85&100.0&0.0&0.0&0.0&0.0&0.0\\
0.5&90&100.0&0.0&0.0&0.0&0.0&0.0\\
0.5&95&100.0&0.0&0.0&0.0&0.0&0.0\\
0.5&100&100.0&0.0&0.0&0.0&0.0&0.0\\
0.5&105&100.0&0.0&0.0&0.0&0.0&0.0\\
0.6&75&0.0&0.0&34.0&51.0&8.0&7.0\\
0.6&80&0.0&0.0&29.0&50.0&8.0&13.0\\
0.6&85&0.0&0.0&29.0&57.0&10.0&4.0\\
0.6&90&100.0&0.0&0.0&0.0&0.0&0.0\\
0.6&95&100.0&0.0&0.0&0.0&0.0&0.0\\
0.6&100&100.0&0.0&0.0&0.0&0.0&0.0\\
0.6&105&100.0&0.0&0.0&0.0&0.0&0.0\\
0.7&75&0.0&1.0&57.0&32.0&2.0&8.0\\
0.7&80&0.0&2.0&33.0&48.0&10.0&8.0\\
0.7&85&0.0&0.0&37.0&46.0&11.0&6.0\\
0.7&90&0.0&0.0&44.0&37.0&12.0&7.0\\
0.7&95&0.0&0.0&40.0&43.0&5.0&12.0\\
0.7&100&100.0&0.0&0.0&0.0&0.0&0.0\\
0.7&105&100.0&0.0&0.0&0.0&0.0&0.0\\
0.8&75&0.0&2.0&66.0&29.0&0.0&4.0\\
0.8&80&0.0&9.0&51.0&38.0&0.0&4.0\\
0.8&85&0.0&6.0&50.0&36.0&2.0&6.0\\
0.8&90&1.0&2.0&41.0&42.0&7.0&7.0\\
0.8&95&0.0&0.0&44.0&40.0&7.0&9.0\\
0.8&100&0.0&0.0&69.0&17.0&7.0&7.0\\
0.8&105&50.0&0.0&33.0&9.0&4.0&4.0\\
0.9&75&0.0&1.0&69.0&23.0&0.0&8.0\\
0.9&80&0.0&0.0&77.0&17.0&2.0&4.0\\
0.9&85&0.0&14.0&56.0&26.0&1.0&6.0\\
0.9&90&0.0&10.0&63.0&23.0&1.0&4.0\\
0.9&95&1.0&14.0&48.0&32.0&2.0&3.0\\
0.9&100&0.0&3.0&65.0&24.0&5.0&3.0\\
0.9&105&0.0&0.0&67.0&17.0&7.0&9.0\\
1.0&75&0.0&1.0&71.0&22.0&0.0&6.0\\
1.0&80&0.0&6.0&71.0&21.0&0.0&3.0\\
1.0&85&0.0&3.0&75.0&20.0&0.0&3.0\\
1.0&90&0.0&18.0&68.0&14.0&0.0&0.0\\
1.0&95&0.0&20.0&62.0&11.0&2.0&5.0\\
1.0&100&0.0&15.0&64.0&17.0&3.0&1.0\\
1.0&105&3.0&10.0&59.0&21.0&3.0&5.0\\
1.1&75&0.0&0.0&74.0&23.0&0.0&3.0\\
1.1&80&0.0&0.0&85.0&15.0&0.0&0.0\\
1.1&85&0.0&4.0&80.0&16.0&0.0&2.0\\
1.1&90&0.0&12.0&81.0&7.0&0.0&0.0\\
1.1&95&0.0&14.0&73.0&11.0&1.0&1.0\\
1.1&100&0.0&23.0&58.0&17.0&1.0&4.0\\
1.1&105&1.0&18.0&62.0&18.0&0.0&1.0\\
1.3&75&0.0&0.0&77.0&20.0&0.0&3.0\\
1.3&80&0.0&2.0&82.0&13.0&0.0&3.0\\
1.3&85&0.0&5.0&83.0&10.0&0.0&2.0\\
1.3&90&0.0&4.0&87.0&9.0&0.0&0.0\\
1.3&95&0.0&13.0&78.0&9.0&0.0&0.0\\
1.3&100&0.0&18.0&74.0&7.0&1.0&0.0\\
1.3&105&1.0&26.0&66.0&7.0&0.0&2.0\\
1.5&75&0.0&0.0&80.0&17.0&0.0&3.0\\
1.5&80&0.0&0.0&91.0&5.0&0.0&4.0\\
1.5&85&0.0&1.0&92.0&7.0&0.0&0.0\\
1.5&90&0.0&7.0&88.0&5.0&0.0&1.0\\
1.5&95&0.0&16.0&71.0&13.0&0.0&1.0\\
1.5&100&0.0&15.0&83.0&2.0&0.0&0.0\\
1.5&105&0.0&25.0&73.0&2.0&0.0&1.0\\
1.7&75&0.0&0.0&89.0&6.0&0.0&5.0\\
1.7&80&0.0&1.0&93.0&5.0&0.0&1.0\\
1.7&85&0.0&0.0&96.0&4.0&0.0&0.0\\
1.7&90&0.0&1.0&92.0&7.0&0.0&0.0\\
1.7&95&0.0&6.0&89.0&5.0&0.0&1.0\\
1.7&100&0.0&17.0&77.0&6.0&0.0&1.0\\
1.7&105&0.0&14.0&83.0&3.0&0.0&0.0\\

\label{tab-res}
\end{deluxetable}

\end{document}